\pdfoutput=1
\documentclass[twocolumn]{aastex62}

\usepackage{graphicx}

\def\kms{${\rm km~s}^{-1}$}

\received{2018 June 19}
\revised{2018 August 2}
\accepted{2018 August 2}
\published{2018 August 17}
\shorttitle{Witnessing galaxy assembly at the edge of the reionization epoch}
\shortauthors{V. D'Odorico et al.}
\begin{document}

\title{Witnessing galaxy assembly at the edge of the reionization epoch \footnote{Based on ALMA and ESO VLT observations.}}

\correspondingauthor{V. D'Odorico}
\email{valentina.dodorico@inaf.it}

\author[0000-0003-3693-3091]{V. D'Odorico}
\affiliation{INAF Osservatorio Astronomico di Trieste, via G. Tiepolo 11, I-34143 Trieste, Italy }
\affiliation{Scuola Normale Superiore, Piazza dei Cavalieri 7, I-56126, Pisa, Italy}
\author[0000-0002-4227-6035]{C. Feruglio}
\affil{INAF Osservatorio Astronomico di Trieste, via G. Tiepolo 11, Trieste, Italy }
\author[0000-0002-9400-7312]{A. Ferrara}
\affiliation{Scuola Normale Superiore, Piazza dei Cavalieri 7, 56126, Pisa, Italy}
\author[0000-0002-7200-8293]{S. Gallerani}
\affiliation{Scuola Normale Superiore, Piazza dei Cavalieri 7, 56126, Pisa, Italy}
\author[0000-0002-7129-5761]{A. Pallottini}
\affiliation{Scuola Normale Superiore, Piazza dei Cavalieri 7, 56126, Pisa, Italy}
\affiliation{Centro Fermi, Museo Storico della Fisica e Centro Studi e Ricerche ``Enrico Fermi'', Piazza del Viminale 1, I-00184 Roma, Italy}
\author[0000-0002-6719-380X]{S. Carniani}
\affiliation{Cavendish Laboratory, University of Cambridge, 19 J. J. Thomson Avenue, Cambridge CB3 0HE, UK}
\affiliation{Kavli Institute for Cosmology, University of Cambridge, Madingley Road, Cambridge CB3 0HA, UK}
\author{R. Maiolino}
\affiliation{Cavendish Laboratory, University of Cambridge, 19 J. J. Thomson Avenue, Cambridge CB3 0HE, UK}
\affiliation{Kavli Institute for Cosmology, University of Cambridge, Madingley Road, Cambridge CB3 0HA, UK}
\author[0000-0002-2115-5234]{S. Cristiani}
\affiliation{INAF Osservatorio Astronomico di Trieste, via G. Tiepolo 11, I-34143 Trieste, Italy }
\author[0000-0002-9889-4238]{A. Marconi}
\affiliation{Dipartimento di Fisica e Astronomia, Universita di Firenze, via G. Sansone 1, I-50019, Sesto Fiorentino (Firenze), Italy}
\affiliation{INAF-Osservatorio Astrosico di Arcetri, Largo E. Fermi 2, I-50125, Firenze, Italy}
\author[0000-0001-9095-2782]{E. Piconcelli}
\affiliation{ INAF Osservatorio Astronomico di Roma, via Frascati 33, I-00078 Monte Porzio Catone, Italy}
\author{F. Fiore}
\affiliation{INAF Osservatorio Astronomico di Trieste, via G. Tiepolo 11, I-34143Trieste, Italy }



\begin{abstract}
We report the discovery of Serenity-18, a galaxy at $z\simeq5.939$ for which we could measure the content of molecular gas, $M({\rm H}_2)\simeq 5 \times10^9$ M$_{\sun}$, traced by the CO(6-5) emission, together with the metal poor ([Fe/H]~$=-3.08 \pm 0.12$, [Si/H]~$=-2.86\pm0.14$) gas clump/filament which is possibly feeding its growth. The galaxy has an estimated star formation rate of $\approx100$ $M_{\sun}$ yr$^{-1}$, implying that it is a typical main sequence galaxy at these redshifts. The  metal poor gas  is detected through a damped Lyman-$\alpha$ absorber (DLA) observed at a spatial separation of 40 kpc and at the same redshift of Serenity-18,  along the line of sight to the quasar SDSS J2310+1855 ($z_{\rm em}\simeq6.0025$). The chemical abundances measured for the damped Lyman-$\alpha$ system are in very good agreement with those measured for other DLAs discovered at similar redshifts, indicating an enrichment due to massive PopII stars. The galaxy/damped system that we discovered is a direct observational evidence of the assembly of a galaxy 
%through accretion from the inter-galactic medium 
at the edge of the reionization epoch.
  \end{abstract}



\section{Introduction} 
\label{sec:intro}

Measuring the molecular gas content in early galaxies ($z\gtrsim 6$) is fundamental to reconstruct the cosmic star formation history, reionization, and enrichment. Molecular hydrogen is typically traced by emission from the carbon monoxide molecule, CO, because H$_2$ itself is generally not observable.
While for a few galaxies at $z\approx6-8$ -- less than 1 billion years from the Big Bang -- we have sparse data on the stellar content (e.g. Bouwens et al. 2015; Jiang et al. 2016), the diffuse atomic gas (e.g. Carniani et al. 2017, 2018a) and the dust amount (e.g. Laporte et al. 2017; Hashimoto et al. 2018; Tamura et al. 2018), we completely lack information on their dense molecular component.  
At $z\sim6$, CO has been detected only in a few bright quasars tracing massive, highly star-bursting galaxies (Wang et al 2010, 2013; Gallerani et al. 2014; Venemans et al. 2017; Feruglio et al. 2018), which are rare objects that do not represent the bulk of the galaxy population in the early universe (e.g. Robertson et al. 2015). 

At lower redshift ($z\lesssim 4$), cold gas emission from galaxies has been recently detected in association with metal rich   damped Lyman-$\alpha$ absorbers (DLAs), the strongest \ion{H}{1} absorptions observed in quasar  spectra (characterized by column densities $N($\ion{H}{1}$) \ge 2 \times 10^{20}$ cm$^{-2}$).  DLAs have the advantage that their detection is not biased by luminosity; on the other hand, the direct identification of their host galaxies has proven to be  extremely challenging, in particular at high redshift (e.g. Fumagalli et al. 2017).

Thanks to the lower luminosity of the quasar at sub-mm wavelengths, it was possible to observe with  Atacama Large Millimeter/submillimeter Array (ALMA) four high-redshift ($z\sim2-4$), high-metallicity ($Z \sim0.1-1.0$ $Z_{\sun}$) DLAs revealing cold gas emission, either [\ion{C}{2}] or CO, in associated galaxies (Neeleman et al. 2017, 2018; Fynbo et al. 2018). The impact parameter between the galaxy detected in emission and the DLA varies between 18 and 117 kpc, implying that the DLA could arise in the circumgalactic medium of the target galaxy, but also that it could be associated with  a fainter, undetected object. 

%The identification of the host, however, demonstrates to be extremely challenging (Fumagalli et al. 2017).  
%The nature of galaxies associated with DLA remains therefore uncertain, expecially at the highest redshifts.  
%The most efficient way to detect emission from stars and ionised gas in the galaxies associated with low redshift DLA 
%are the new-generation IFU, operating at optical wavelengths (Peroux et al. 2011, Bouch\'e et al. 213, Fumagalli et al. 2017). 
%These observations, however, have the obvious shortcoming of being significantly affected by obscuration by dust. 
%The cold gas phase in DLA host  galaxies can be probed today with ALMA. 
%Observations at sub-mm wavelengths with ALMA have the advantage of being less affected by dust obscuration. 
%Recently,  Neeleman et al. (2017, 2018) provided the first detection of cold gas emission, either [CII] or CO at high $z$, from DLA host galaxies %at $z\sim4$ and 2.2, respectively, showing that these systems are gas rich. 

In this Letter we report the serendipitous discovery of a typical main sequence galaxy at $z\simeq 5.939$ (that we dubbed Serenity-18) detected in CO, associated with a metal poor  DLA observed along the sightline to the quasar J231038.88+185519.7 ($z_{\rm em}=6.0025$, J2310 hereafter), using ALMA and XSHOOTER/Very Large Telescope (VLT)  observations. 
Throughout this Letter we assume a standard flat $\Lambda$CDM cosmology, with $\Omega_\Lambda=0.7$ and $H_0=70$ \kms. 

\section{Observations}

%\subsection{X-Shooter observations}

%We have used X-SHOOTER at the VLT (Vernet et al. 2011) archival data  to investigate the optical/NIR spectrum of the object.
%There are two frames available in the Archive with exposure time 1200 s each, observed with a slit
%of $0.9$ arcsec and a binning $2\times2$ in the VIS arm and a slit of
%$0.6$ arcsec in the NIR arm. The chosen slits corresponds to nominal resolving
%powers $R\simeq 8800$ and  $8100$, respectively.

%The spectra have been reduced with the ESO pipeline (Modigliani et
%al. 2010) with a
%manual localisation of the object and adopting the sky subtraction
%method BSPLINE1 in the VIS, and MEDIAN in the NIR. The 1D, flux
%calibrated spectra produced by the pipeline were then correct for
%telluric absorptions using the ESO tool {\tt Molecfit} (Smette et al. 2015; Kausch et al. 2015). 
%The final spectra,
%obtained combining the two frames, 
%were rebinned to a step of 0.4 and 0.6 \AA, respectively.

%\subsection{ALMA observations}

We have observed the field of the quasar J2310, at $z_{\rm em}=6.0025$ with ALMA band 3 receivers tuned to cover the frequency ranges $[84.56-87.94]$, and $[96.56-99.69]$ GHz (results about the quasar host galaxy are presented in Feruglio et al. 2018). 
Spectral window 1 was tuned at the expected redshifted frequency of CO(6-5) of the quasar J2310, i.e. at 98.75 GHz. 
We performed calibration in the CASA environment (McMullin et al. 2007).
Mapping and data analysis were performed both in the CASA and in the GILDAS (Guilloteau et al. 2000) 
environments.
% (the latter after converting CASA into GILDAS visibility tables). 
By adopting a natural weighting scheme with detection threshold 0.5 times the noise per channel, we obtain a synthesized beam of $0.6 \times 0.4$ arcsec$^2$ at a PA~$=-6$ deg. 
The noise levels are 5.4 $\mu$Jy/beam in the continuum in the aggregated bandwidth, and 0.13 mJy/beam in 23.7 \kms\ wide channels 
(i.e. the maximum spectral resolution of our data).
%By adopting a Briggs algorithm, the synthetic beam is $0.51 \times 0.28$ arcsec$^2$, at a PA$=-11$ deg, 
%and the $1\sigma$ noise level is 0.15 mJy/beam in 23.7 \kms\ wide channels. 
%The results from natural weighting cleaning will be used in the following.

 We also make use of observations obtained with the ALMA 12 m array in Cycle 3, 
 %project 2015.1.00997.S between January and June 2016, 
 covering the frequency ranges [254--257.65] GHz and [269.65--273.15] GHz. 
The data were calibrated and imaged in CASA v4.7 by applying a natural weighting. 
The $1\sigma$ r.m.s sensitivity is 50 $\mu$Jy/beam in the continuum, and 0.20 mJy/beam per 100 \kms\ channel in the spectral data. 
The synthesized beam is $0.9\times0.6$ arcsec$^2$ at a PA~$=49$ deg.
All data were corrected for the primary beam efficiency.

Quasar J2310 was also observed  with XSHOOTER with a nominal resolving power of $R\simeq 8800$ and  $8100$, in the visible (VIS) and near-infrared (NIR) arm, respectively. 
The characteristics of the spectrum and the reduction process were described in Feruglio et al. (2018). We measure an average signal-to-noise ratio (S/N) of $\approx 30$ per resolution element in the region between the Lyman-$\alpha$ (Ly$\alpha$, hereafter) emission and $\sim1$ $\mu$m. In the NIR regions free from sky emission lines and strong telluric absorptions, the S/N is $\approx 15$ per resolution element between 1 and 1.3  $\mu$m and it increases to $\approx 28$ between 1.6 and 1.8 $\mu$m.

\section{Results}

\subsection{A proximate DLA at $z\approx5.939$}
%\subsection{Identified absorption systems}

We inspected the XSHOOTER spectrum by eye, to detect prominent absorption systems. Absorption lines 
were identified and then fit with Voigt profiles using the context LYMAN of the ESO MIDAS software package
(Fontana \& Ballester 1995). 

%The reported redshift is the one of  the strongest component of the velocity profile. 
 
%We identified three \ion{Mg}{2} doublets at $z=2.10468\pm0.00002$, $z=2.242879\pm0.000003$ and $z=2.351088\pm0.00008$, 
%the latter showing also the associated transitions of \ion{Fe}{2} $\lambda\lambda\, 2586, 2600$ \AA.
%
%\item $z=2.10468\pm0.00002$.  Possible weak Mg~II doublet with no other
%associated transition.
%\par \noindent
%\item $z=2.242879\pm0.000003$, Strong Mg~II doublet with a velocity profile
%extended over more than $\sim220$ km/s and 4 observed components. No
%other transitions have been detected.
%\par\noindent
%\item $z=2.351088\pm0.00008$. Mg~II doublet characterised by a weak
%components and a strong one. At the redshift of the latter we
%identified also the transitions of Fe~II $\lambda\lambda\, 2586,
%2600$ \AA.
%\par \noindent
%
%A strong \ion{Fe}{2} multiplet is found at  $z=4.01373\pm0.00005$, showing the transitions at 
% $\lambda\lambda\, 2344, 2382, 2586$ and 2600 \AA, the associated\ion{ Al}{3} doublet 
% $\lambda\lambda\, 1854, 1862$ \AA\ and  a possible \ion{Si}{2} $\lambda\, 1808$ \AA.
%The Mg~II doublet falls in the telluric trough between the J and H bands. 
%
%Two \ion{C}{4} doublets were identified at $z=4.80985\pm0.00006$ and $z=5.28772\pm0.00005$ with no other
%associated transition.

\begin{figure*}
%\plotone{DLA_fit.pdf}
\includegraphics[width=15cm,height=7cm]{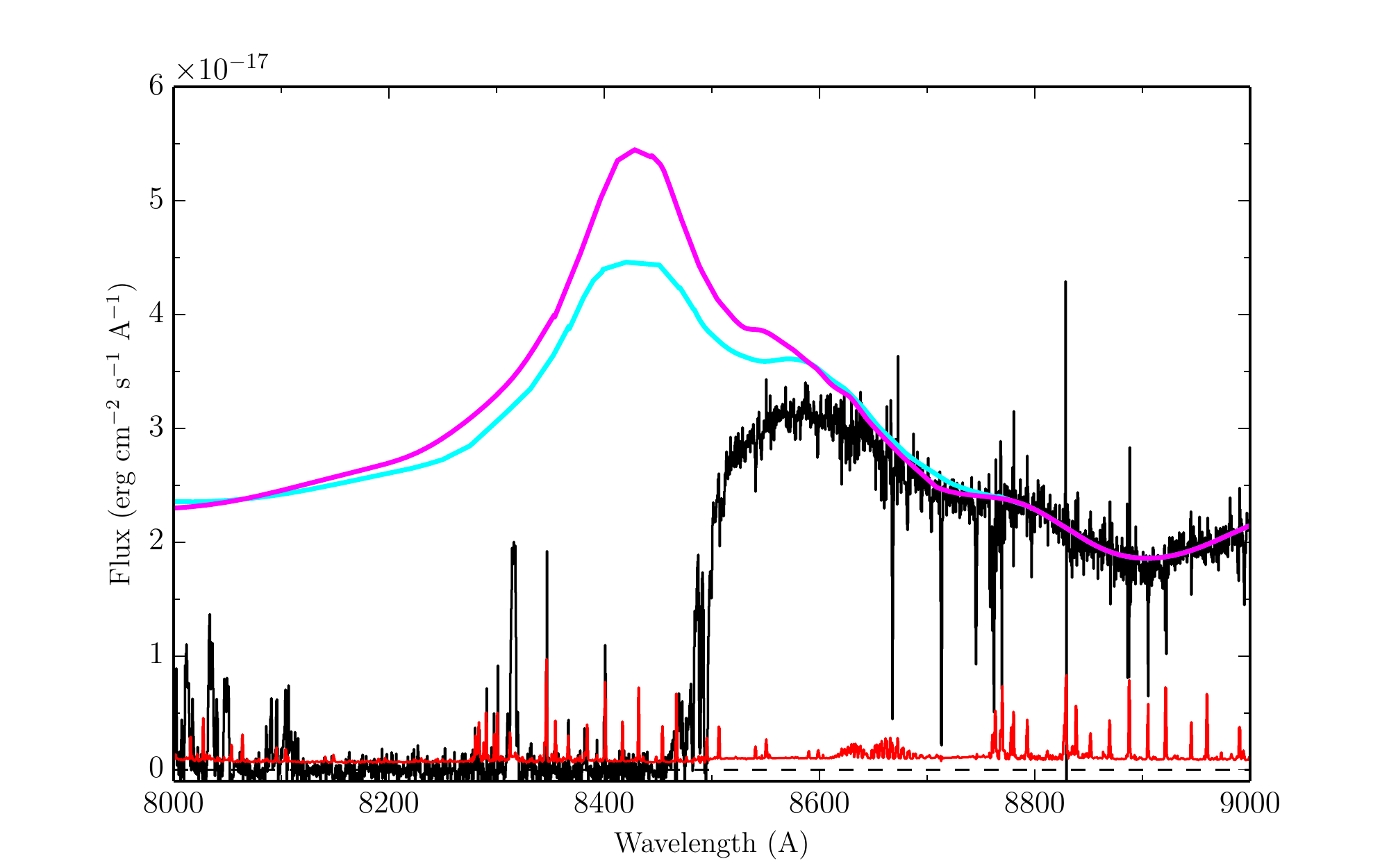}
\includegraphics[width=15cm]{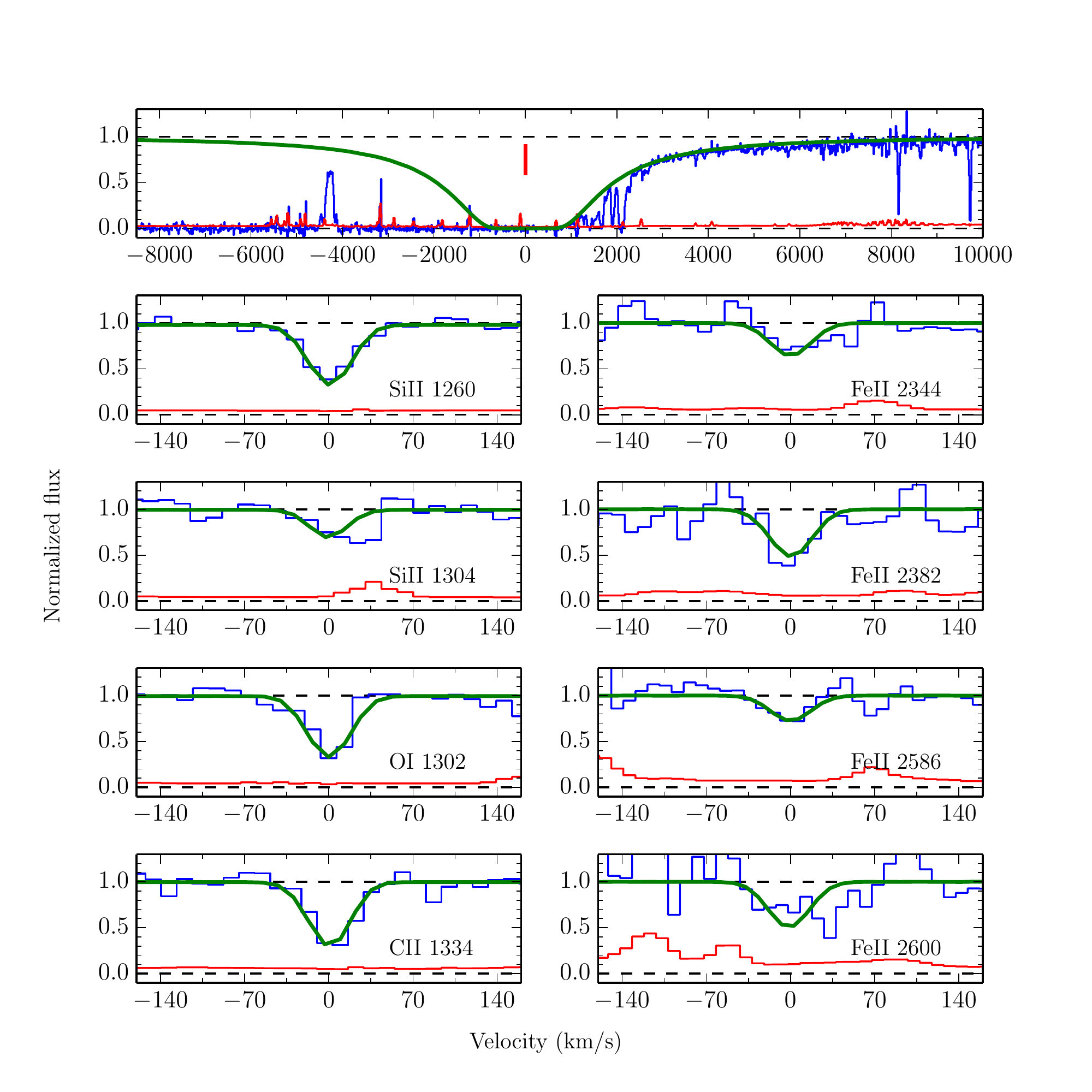}
%\gridline{\fig{DLA_cont.pdf}{0.7\textwidth}{(a)}}
%\gridline{\fig{DLA_fit.pdf}{0.9\textwidth}{(b)}}
%\plottwo{DLA_cont.pdf}{DLA_fit.pdf}
\caption{Upper panel: flux-calibrated spectrum of J2310 in the region of the Ly$\alpha$ emission with the two continua used to normalize it. Lower panels:  fit of the observed transitions for the proximate DLA in the spectrum of J2310. In each panel, the observed spectrum is shown in blue, the error in red, and the Voigt fit in green. The top panel shows the fit of the \ion{H}{1} Ly$\alpha$ velocity profile. The red  mark indicates the redshift $z=5.938646$, which is also the zero-point of the velocity scale in all the other panels. }
\label{Fig:DLA_fit}
\end{figure*}

The most interesting system in the analyzed spectrum\footnote{The other absorption systems detected in the spectrum will 
be described in a subsequent work.} is a low ionization absorber at 
$z=5.938646\pm0.000007$  ($\Delta v \simeq -2746$ km s$^{-1}$ from the quasar emission redshift) for which 
 we detect strong  transitions due to \ion{O}{1} $\lambda\,1302$ \AA, \ion{C}{2} $\lambda\,1334$ \AA, 
\ion{Si}{2} $\lambda\lambda\, 1260, 1304$ \AA\ and \ion{Fe}{2} $\lambda\lambda\,
2344, 2382, 2586, 2600$ \AA\ (Fig.~\ref{Fig:DLA_fit}). 
The absorption line due to \ion{Al}{2} $\lambda\,1670$ \AA\ is severely affected by a sky line. The presence of the likely saturated
\ion{O}{1} absorption suggests that this could be a DLA, because \ion{O}{1} is a tight tracer of neutral hydrogen that has 
a very similar ionization potential.  
There is also a possible detection of the associated weak \ion{C}{4} doublet at a slightly different
redshift, $z= 5.9392\pm0.0001$, corresponding to $\Delta v\simeq 24$
km s$^{-1}$. A shift between low and high ionization transitions is often
observed in damped systems.

\begin{deluxetable}{lccc}
\tablecaption{Properties of the DLA \label{tab:dla}}
\tablecolumns{3}
%\tablenum{2}
\tablewidth{0pt}
\tablehead{
%\colhead{ALMA J231038.44+185521.95}
Ion & $\log N$  &  [X/H]$^{\rm a}$ & [X/Fe] 
}
\startdata
\ion{H}{1} & $21.05\pm0.10$ & - & - \\
\ion{C}{2}$^{\rm b}$ & $14.73\pm0.18$ & $\ge -2.95$ &  $\ge 0.1$ \\
\ion{O}{1}$^{\rm b}$  & $15.00\pm0.12$ & $\ge -2.9$ &$\ge 0.2$ \\
\ion{Si}{2} & $13.70\pm0.09$ & $-2.86\pm0.14$ & $0.22\pm0.12$\\
\ion{Fe}{2} & $13.47\pm0.06$ & $-3.08\pm0.12$ & - \\
\enddata
%\tablenotetext{a}{At exposure start.}
\tablecomments{$^{\rm a}$Solar abundances and corresponding uncertainties from Asplund et al. (2009). $^{\rm b}$Detected absorption lines are possibly saturated. }
\end{deluxetable}

 \begin{figure*}
%\plotone{dla-plot-fit1gauss.pdf}
\includegraphics[width=18cm]{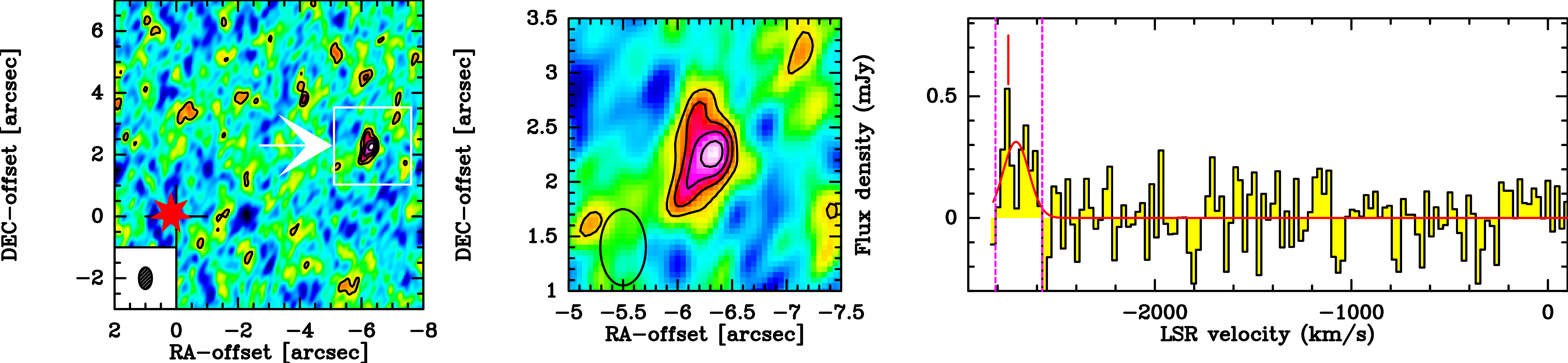}
\caption{Left panel: the velocity-integrated map of the CO(6-5) line of the DLA host galaxy (indicated by a white arrow), located about -6.3, +2.2 arcsec offset from the quasar position (red star). Contours are $2, 3, 4, 5\sigma$, 1$\sigma=5.1~\mu$Jy/beam.
Middle panel: a zoomed view onto the DLA emission. Right panel: the spectrum extracted from the region with $\ge2\sigma$. 
The red curve is the fit to the CO(6-5) emission, while the red vertical mark corresponds to the redshifted frequency of the CO(6-5) line expected at the redshift of the DLA. The dashed magenta lines show the range where the line emission has been integrated to produce the velocity-integrated map (left panel). }
\label{Fig:DLA_emiss}
\end{figure*}

%\begin{figure}
%\plotone{continuum-dla}
%\caption{The 3 mm continuum (430 $\mu$m continuum map obtained by averaging spectral windows 2,3,4.  }
%\end{figure}

\begin{figure*}
%\plotone{DLA_abund.pdf}
\includegraphics[width=9cm]{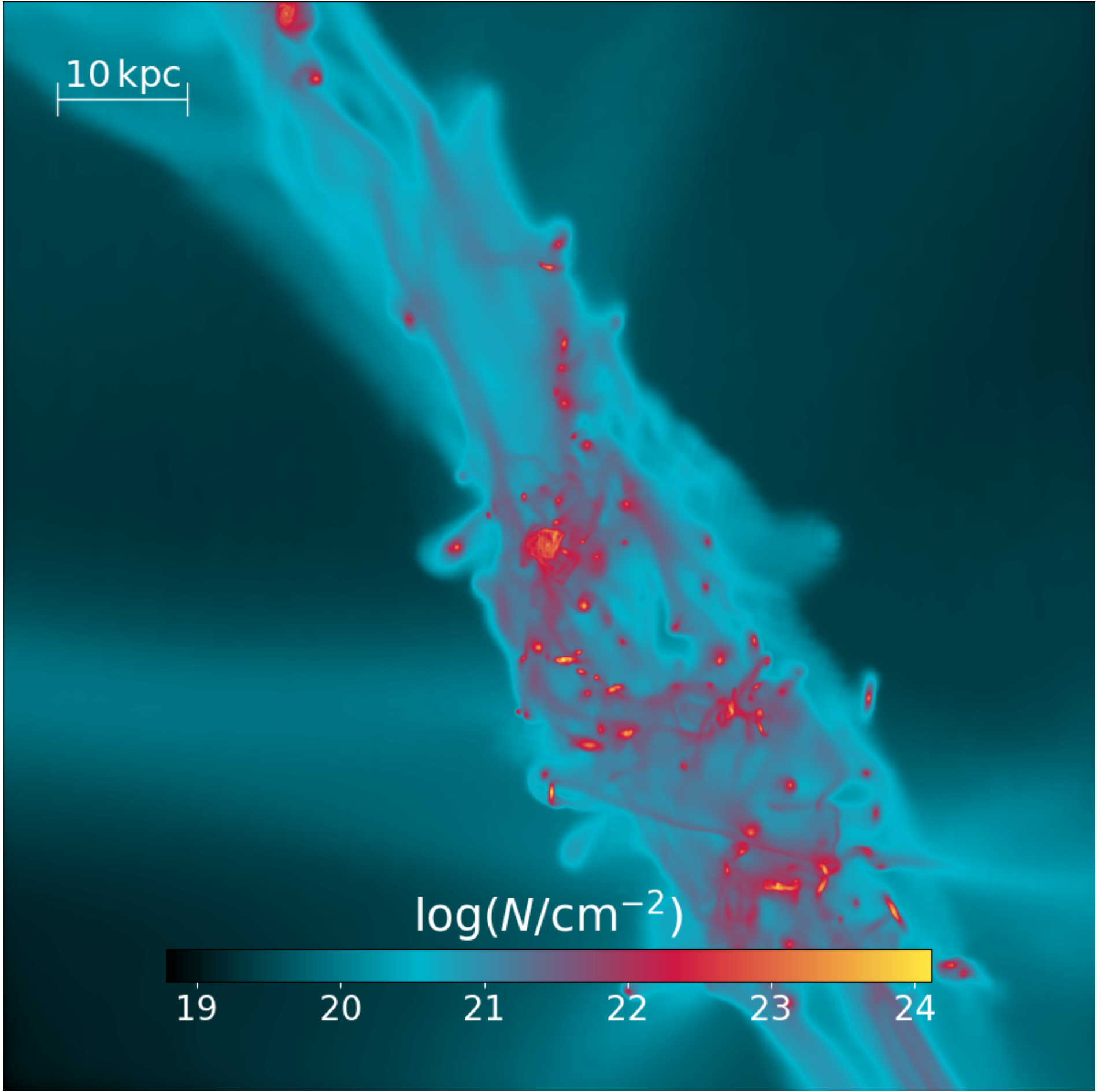}
\includegraphics[width=9cm]{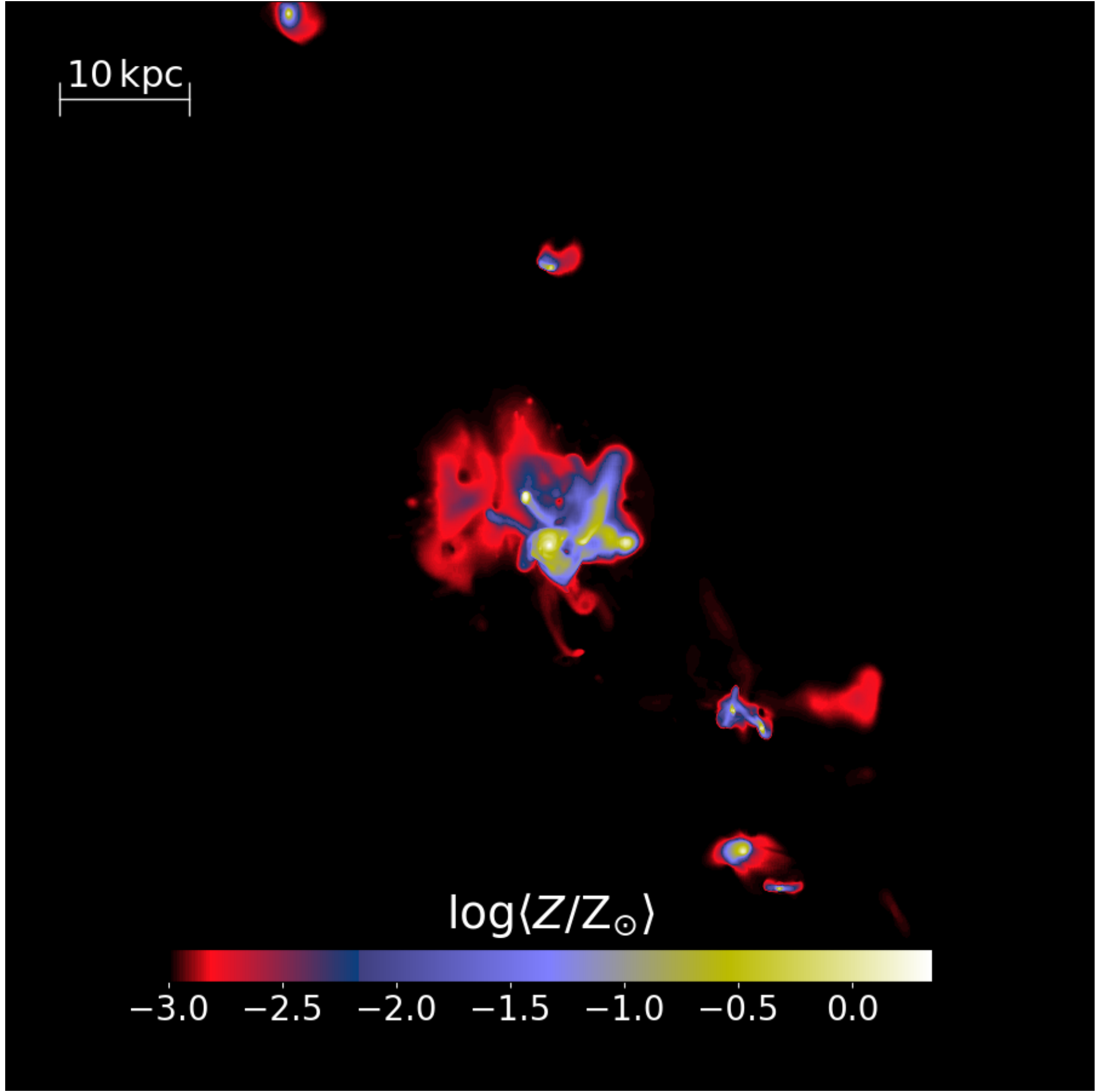}
%\includegraphics[width=16cm]{DLA_abund.pdf}
%\caption{Abundances measured for the DLA J2310 compared with literature data. Red star: the studied DLA; Blue points: high-z DLA from Becker et al. (2012); %Green triangles: metal poor DLAs in the XQ-100 sample (Berg et al. 2016). The dark and light shaded bands shown in the plots represent the standard deviation %and the 95 \% confidence interval of VMP DLAs (Cooke et al. 2011).   }
\caption{Maps of the \ion{H}{1} column density (left panel) and metallicity (right panel) for the simulated  galaxy Alth{\ae}a at $z\simeq6$, which lies at the center of each panel (Pallottini et al. 2017). The maps are 80 kpc on a side. Serenity-18 has characteristics very similar to Alth{\ae}a, in this framework the DLA could trace the gas in the filament or in the periphery of one of the small clumps/satellites that are embedded in the same filament as the galaxy and that will eventually feed its growth.   }
\label{Fig:sim}
\end{figure*}

DLAs at these redshifts generally lack the information on the \ion{H}{1} column density because the 
spectrum blueward of the quasar Ly$\alpha$ emission is completely absorbed. In the present case, the  DLA falls in the region within 5000 km s$^{-1}$ of the quasar systemic redshift (called the ``proximity region''). This allowed us to estimate  the column density, $N$(\ion{H}{1}), by fitting the damping wing   at the position of the Ly$\alpha$ in emission (Fig.~\ref{Fig:DLA_fit}).  
To fit the DLA profile, we estimated the continuum level in the region of the Ly$\alpha$/Ly$\beta$ emissions with the following procedure. We considered 
all of the Principal Component Analysis continua computed for the quasars in the XQ-100 sample (L\'opez et al. 2016). The continua were redshifted to the systemic redshift of J2310 and compared to its XSHOOTER spectrum. Only two continua matched reasonably well the large width of the observed emission lines of J2310 (Fig.~\ref{Fig:DLA_fit}, upper panel). We fit the DLA \ion{H}{1} Ly$\alpha$ and Ly$\beta$ absorption lines using the two continua separately  to estimate a more reliable uncertainty on the measure. The \ion{H}{1} lines were centered at the redshift determined from the fit of the low ionization heavy element transitions. The adopted column density and its error, $\log N($HI$) = 21.05 \pm 0.10$ [cm$^{-2}$], are the average of the two column density measurements and the half width of the interval spanned by the measurements and their $1\sigma$ errors, respectively.  
Figure~\ref{Fig:DLA_fit} reports the results of the fit for all detected transitions. Absorption lines due to \ion{O}{1}, \ion{C}{2} and \ion{Si}{2} were fitted simultaneously adopting the same Doppler parameter, $b=8.0\pm0.4$ km s$^{-1}$. 
This is a reasonable assumption, as metal lines are not resolved in the XSHOOTER spectrum.  Detected lines of the \ion{Fe}{2} multiplet were fitted together (excluding the noisy \ion{Fe}{2} 2600 \AA\ line) resulting in a Doppler parameter $b=10.5\pm1.5$ km s$^{-1}$. 
Column densities for the observed ions and relative chemical abundances are reported in Table~\ref{tab:dla} and discussed in Sec.~4.1. 
 
\begin{deluxetable}{ll}
\tablecaption{Properties of Serenity-18 \label{tab:gal}}
\tablecolumns{2}
%\tablenum{2}
\tablewidth{0pt}
\tablehead{
\colhead{ALMA J231038.44+185521.95}
}
\startdata
R.A. (J2000) & 23:10:38.44  \\
Decl. (J2000) & 18:55:21.95 \\
Redshift of CO(6-5) emission  &  5.93957\\
Impact parameter [arcsec] & 6.7 \\
$\rm FWHM_{CO(6-5)}$ [km s$^{-1}$] &   $155 \pm  30$ \\ 
$\int S_{{\rm CO}(6-5)}dv$ [Jy~ km s$^{-1}$]  &  $0.06\pm0.012$  \\ 
$L^\prime {\rm CO}(6-5)$ [K~km s$^{-1}$~pc$^{-2}$]  &  $(2\pm0.4)\times10^9$ \\
$M({\rm H}_2)~[M_\odot]$ &   $(5.4\pm0.5) \times 10^{9}$ \\
$M_{\rm dyn}sin^2(i)~[M_\odot] $ &  $\leq 5.6\times 10^9$     \\
$L_{\rm FIR} ~ [L_\odot]$ &   $\approx 10^{12}$  \\ 
SFR [$M_\odot {\rm yr}^{-1}$]     &   $\approx 115$   \\
\enddata
%\tablenotetext{a}{At exposure start.}
%\tablecomments{The ``C'' command column identifier in the 3 column turns on
%}
\end{deluxetable}

\subsection{Emission line and sub-mm continuum}

Following the identification of the proximate DLA absorber in the spectrum of J2310, we analyzed our ALMA data and found an emission line at a frequency 99.642 GHz, corresponding to a velocity of $-2710$ km s$^{-1}$ with respect to the quasar host galaxy ($z_{\rm em}=6.0025$). 
The line is located at the very edge of the bandpass, therefore the continuum-level can be estimated only on the right-hand side of the line  (Fig.~\ref{Fig:DLA_emiss}, right panel) and there is the possibility that part of the emission is falling outside the observed bandpass. This implies that the integrated flux of the emission line and its derived quantities might be lower limits. 
A Gaussian fit to the emission line gives an $\rm FWHM=155\pm30$ \kms, and an integrated intensity of $0.053\pm0.01$ Jy km s$^{-1}$.

By integrating over the line width, and collapsing the corresponding spectral channels, we find a 5$\sigma$ emission on the map at a position 
R.A., decl. ($23:10:38.437, 18:55:22.00$), or about ($-6.3,+2.2$) arcsec offset from the quasar position (Fig.~\ref{Fig:DLA_emiss}, left panel). This is well within the primary beam of the ALMA antennas in band 3.
The sky frequency corresponds to the expected frequency of the CO(6-5) emitted at the redshift of the DLA absorption system, therefore it can be identified with the CO(6-5) emission from a galaxy associated with the DLA.
We estimate the probability of a chance coincidence between a foreground CO emitter (CO(3-2) at $z\simeq2.47$ or CO(4-3) at $z\simeq3.63$) and  the DLA absorption, based on the expected number of emission lines in ALMA band 3 with flux similar to the one observed here (Walter et al. 2016). \ 
Assuming that a redshift difference $\Delta v \lesssim \pm100$ \kms,  comparable with the rotation velocity of a disk at these redshifts, is required to associate   the CO emission and the DLA absorption lines, we obtain a probability of chance alignment of $\sim1$ ~\% $-2$~\%.
  
We have also measured the CO line in the Fourier plane, after averaging visibilities over the line width and shifting the corresponding visibility table to the phase center.
% (uv-shift task within GILDAS/MAPPING).   
A fit with a point source gives a zero spacing flux density of $251\pm51\rm ~\mu Jy$, which corresponds to an integrated intensity of 
 Sdv~$= 0.06\pm0.012$ Jy km s$^{-1}$, consistent with that derived from the fit of the spectrum. 
The source  may appear slightly more extended than the synthesized beam on the integrated map (Fig.~\ref{Fig:DLA_emiss}, central panel). To verify this possibility, we also fitted the visibilities with a circular or elliptical gaussian function, but the fit does not converge. We conclude that the source is not resolved in our data. 
 
The galaxy is not detected in the 3 mm continuum, down to a 3$\sigma$ upper limit of 16.2 $\mu$Jy/beam. 

We found no detection in the rest-frame 150 $\mu$m continuum probed by band 6 ALMA data, down to a 3$\sigma$ upper limit of 150 $\mu$Jy/beam.
For a typical spectral energy distribution (SED) of a star-forming galaxy with a dust temperature in the range $ T_{dust}=30-50~ K$, and located at this redshift, this limit correponds to a  total infrared luminosity $L({\rm FIR})\approx 1.2 \times 10^{12}~ L_{\odot}$.
We note that the bandpass does not include the expected frequencies of other far-infrared (FIR) emission lines, included the brightest one, namely [CII].

The properties of Serenity-18 are summarized in Table~\ref{tab:gal}.

\section{Discussion}

\subsection{DLA properties}
On the basis of the large \ion{H}{1} column density measured for the DLA system, we assumed that no ionization corrections were needed and carried out the computation of the relative chemical abundances based on the column densities  and the solar abundances from Asplund et al. (2009). 
The iron and silicon abundances, [Fe/H]~$=-3.08 \pm 0.12$ and [Si/H]~$=-2.86\pm0.14$, place this DLA absorber in the very metal poor regime as defined by Cooke et al. (2011). 
The absorption lines due to \ion{C}{2} $\lambda\,1334$ \AA\ and \ion{O}{1} $\lambda\,1302$ \AA\ could be saturated, thus we could derive only lower limits to the abundances of C and O.
Previous studies of DLAs (e.g. Vladilo et al. 2011; Rafelski et al. 2012) have shown that below metallicities  [Fe/H]~$\approx -2$ dust corrections are negligible; as a consequence, we are computing abundances assuming a dust-free gas.  

%In Fig.~\ref{Fig:DLA_abund}, we compare the measured abundances with two main samples: the $z\approx6$ DLAs by Becker et al. (2012; blue points) which %had UVES/VLT observations  and the DLAs identified in the XSHOOTER XQ-100 survey  (Berg et al. 2016; green triangles) from which the systems with %metallicity [M/H]~$\le -1.5$ were selected.   
The studied absorber has abundances in very good agreement with those measured for the sample of $z\approx6$ DLAs by Becker et al. (2012).
%, which had UVES/VLT observations.  
%, with the exception of the [Fe/H] abundance (upper left panel). This  measurement, when possible, is however very uncertain at $z\sim6$. 
%Comparing with the lower redshift XSHOOTER sample, it is important to notice that all systems have larger metallicities, with only 5 systems in the range $-3.0 %<$~[Fe/H]~$\le -2.5$. For the other relative abundances, both limits and measurements are in agreement with what measured at larger redshift. 
%In the plots, we report also the standard deviation and 95 \% confidence interval of the sample of very metal poor (VMP) DLAs selected by Cooke et al. (2011) %with the exclusion of the Carbon enhanced DLA along the line of sight to SDSS J0035-0918. 

As already discussed in Becker et al. (2012), the chemical abundances observed for our system are also consistent with the 95 \% confidence interval of the abundances for the sample of very metal poor DLAs selected by Cooke et al. (2011)  at $z\sim 2-4$\footnote{With the exclusion of the Carbon enhanced DLA along the line of sight to SDSS J0035-0918}. The observed abundance pattern is well explained by the predictions for PopII progenitors with $M \sim20$ $M_{\sun}$. 
%The very low metallicity indicates that the enrichment could be due to a single or very few core-collapse SuperNovae. 

The column density of \ion{H}{1} requires self-shielding of the gas from the cosmic ultraviolet (UV) background, which in turn implies that the absorbing gas density should be larger than $\simeq 0.1$ cm$^{-3}$ (Rahmati et al. 2013); this limit translates into a limit on the physical size of $\lesssim 4$ kpc.  
Together with the low metallicity, this suggests that we are seeing a gas filament/clump that has been recently forming from the intergalactic medium.

\subsection{Serenity-18 properties}

The CO-emitting galaxy is unresolved in our data. We estimate its upper limits size as $D \leq {\rm FWHM}_{\rm beam}=0.6$ arcsec, which corresponds to $\approx3.6$ kpc at the redshift of the galaxy. 
The inclination on the line of sight cannot be estimated either, because the source is unresolved. 
%The narrow FWHM of the line, however, suggests that the galaxy is seen nearly face-on. 
We derive the dynamical mass, modulus the inclination, by applying the relation, 
 $M_{\rm dyn} \sin^2(i) = 1.16~ 10^5 \times (0.75\times $FWHM$_{\rm CO})^2 \times D$ (Wang et al. 2013, Feruglio et al. 2018), where $\rm FWHM_{\rm CO}=155$ \kms, and $D$ is the source size in kpc (diameter). 
From this relation we find, $M_{\rm dyn} \sin^2(i) \leq 5.6~10^9 ~ M_{\odot}$. 

In order to estimate the molecular gas mass from CO(6-5) we need to make some assumptions about the $S_{\rm CO}(6-5)/(1-0)$ ratio, and on the luminosity-to-mass conversion factor. 
We adopt in the following, $S_{\rm CO}(6-5)/(1-0)=r_{61}=20$, which is the average value measured for star forming galaxies from $z=0$ to 4 (Carilli \& Walter 2013). We note that an indication of a higher excitation at $z\sim6$, $r_{61}=70-150$, comes from recent hydrodynamical simulations by Vallini et al. (2018). 

Concerning the $\alpha_{\rm CO}$, the Milky Way value (i.e.  4.3 K \kms\ pc$^{-2}$ M$_{\sun}^{-1}$) is probably unphysical for high-$z$ galaxies.  We therefore adopt a lower value based on Vallini et al. (2018), $\alpha_{\rm CO}=1.5$ K \kms\ pc$^{-2}$ M$_{\sun}^{-1}$. 
Based on these assumptions we infer a molecular gas mass of $M({\rm H}_2)= (5.4\pm0.5) \times 10^{9} \times (\alpha_{\rm CO}/1.5)~ M_{\sun}$.
For different excitation properties of the gas (e.g. Vallini et al. 2018) the mass budget has to be decreased accordingly. 
For an inclination $i=90$ deg (edge-on) and $50$ deg, the molecular gas fraction in the galaxy would be 
$\mu= M({\rm H}_2)/M_{\rm dyn}\sim 0.9$ and 0.6, respectively.

We note that the FWHM estimated by fitting a 1D Gaussian profile to the CO line, FWHM~$=155\pm30$ km s$^{-1}$, is similar to those estimated for [\ion{C}{2}] in typical galaxies at $z \approx 6 - 7$ (FWHM~$\approx150$ km s$^{-1}$, Maiolino et al. 2015; Knudsen et al. 2016; Pentericci et al. 2016; Bradac et al. 2017,  Matthee et al. 2017; Carniani et al. 2018a,b). 
We cannot exclude, however, that part of the line may have been missed out of the bandpass (see Fig.~\ref{Fig:DLA_emiss}). In this case, our determinations of the gas mass and dynamical mass would be lower limits. 

From the calibration of Greve et al. (2014) between $L_{\rm FIR}$ and $L^{\prime}$ CO(6-5), we derive a far-infrared luminosity of $L_{\rm FIR} \approx 10^{12}~L_{\sun}$. 
This is a very rough estimate with an uncertainty of $\sim1$ dex due to the scatter of the correlation. 
We convert $L_{\rm FIR} $ to a star formation rate (SFR) using the Kennicutt et al. (1998) conversion factor corrected for a Kroupa (2011) initial mass function. 
We find  $\rm SFR\approx115~ M_{\odot}/yr$. 
This value is consistent with the upper limit on the 150 $\mu$m continuum estimated from band 6 data. In fact, by assuming a typical SED of a star-forming galaxy with dust temperature in the range $\rm T_{dust}\sim 30-50~K$, we would expect a $3\sigma$ detection of the continuum for a $\rm SFR=200 ~M_{\odot}~yr^{-1}$ (see also Pallottini et al. 2017, Behrens et al. 2018).

The properties we have derived for Serenity-18 allows us to state that this is the first detection of molecular gas emission from a typical main sequence galaxy at the end of the reionization epoch (see e.g. Santini et al. 2017).

\subsection{The system Serenity-18 + DLA}

The difference in redshift between the absorbing gas of the DLA and the emission line is of $\sim50$ \kms.
Their angular separation is of $6.7$ arcsec, corresponding to an impact parameter of $\approx 40$ kpc at the DLA redshift, in a configuration similar to previous results for lower redshift DLAs (e.g. Neeleman et al. 2017, 2018). 
This confirms the association between the CO-emitting galaxy and the  metal poor absorber.  Furthermore, the relatively small redshift difference between the galaxy/DLA system and the quasar ($\Delta v \simeq -2746$ km s$^{-1}$ or $\approx 4$ proper Mpc) suggests that they could all be part of the same large-scale structure. 

State-of-the-art cosmological hydrodynamical simulations can help visualize the overall picture, which could give rise to our observations. We note that the luminosity of the CO(6-5) line, the inferred $M_{\rm dyn}$ and SFR of Serenity-18 are consistent with the predictions for the simulated $z\simeq6$ galaxy Alth{\ae}a (Pallottini et al. 2017; Vallini et al. 2018). In Fig.~\ref{Fig:sim}, we show the \ion{H}{1} column density and metallicity maps for the filament embedding Alth{\ae}a. 
%at $z\simeq6$ (Pallottini et al. 2017), which has very similar characteristics as our galaxy, Serenity-18. 
In this context, the DLA can be identified with either one of the satellites of Alth{\ae}a, or with a gas condensation/filament surrounding the galaxy and possibly feeding it with fresh fuel for star formation.
%Such a configuration is indeed predicted by state-of-the-art cosmological hydrodynamical simulations of early galaxy formation (e.g. Dekel et al. 2009; Pallottini et al. 2017) and
This picture is also in agreement  with observations of clumpy galaxy assembly at $z\gtrsim6$ (Ouchi et al. 2010; Jiang et al. 2013; Bowler et al. 2017; Matthee et al. 2017;  Carniani et al. 2018a, 2018b).

The Serenity-18/DLA complex opens a new window in the study of typical galaxies in the early universe. It represents an ideal target for deeper, multiwavelength (UV/optical/NIR/mm) observations both in imaging and spectroscopy to understand the process of galaxy assembly. It suggests that also metal poor DLAs could be associated with CO-emitting galaxies: this could be true only at the highest redshifts, but it should be tested also at the lower redshifts where it is easier to select DLAs by metallicity.

\acknowledgments
We thank the referee for a careful review. 
This Letter makes use of the following ALMA data: ADS/JAO.ALMA \#2015.1.00584.S and  2015.1.00997.S. ALMA is a partnership of ESO (representing its member states), NSF (USA) and NINS (Japan), together with NRC (Canada), MOST and ASIAA (Taiwan), and KASI (Republic of Korea), in cooperation with the Republic of Chile. The Joint ALMA Observatory is operated by ESO, AUI/NRAO and NAOJ.
Based on observations made with ESO Telescopes at the La Silla Paranal Observatory under programme ID 098.B-0537(A).
We are grateful to Paolo Ventura, Paolo Molaro and Leslie Sage for useful discussions. 
C.F. acknowledges support from  the European Union Horizon 2020 Research and Innovation Programme under the Marie Sklodowska-Curie grant agreement No. 664931. 
A.F. acknowledges support from the ERC Advanced Grant INTERSTELLAR H2020/740120. This research was supported by the Munich Institute for Astro- and Particle Physics (MIAPP) of the DFG cluster of excellence ``Origin and Structure of the Universe''.
R.M. and S.Ca. acknowledge support by the Science and Technology Facilities Council (STFC). R.M. and S.Ca. acknowledges ERC Advanced Grant 695671 {\it QUENCH}. 
F.F. acknowledges financial support from INAF under the contract PRIN INAF 2016 FORECAST.

\end{document}